\documentclass{aastex}
\usepackage{spr-astr-addons}
\usepackage{url}

\RequirePackage{color}

\newcommand{\emaila}{authors@email.com}

\begin{document}

\title{Tracing the evolution of nearby early-type galaxies in low density environments. 
The Ultraviolet view from {\it GALEX}}
\slugcomment{Not to appear in Nonlearned J., 45.}
\shorttitle{Tracing the evolution of nearby early-type galaxies}
\shortauthors{Rampazzo et al.}

\author{R. Rampazzo, F. Annibali} 
\affil{INAF-Osservatorio Astronomico di Padova \\
        Vicolo dell'Osservatorio 5
        35122 Padova, Italy }
\email{\emaila roberto.rampazzo@oapd.inaf.it}
\and \author{A. Marino, L. Bianchi}
\affil{Dept. of Physics and Astronomy \\
        Johns Hopkins University\\
         3400 North Charles Street\\
         Baltimore, MD 21218  USA}
\and \author{ A. Bressan, L.M. Buson, M. Clemens}
\affil{INAF-Osservatorio Astronomico di Padova \\
        Vicolo dell'Osservatorio 5
        35122 Padova, Italy }
\and \author{P. Panuzzo\altaffilmark{}}
\affil{DSM/Irfu/Service d'Astrophysique\\
         CEA Saclay,  91191 Gif sur Yvette Cedex, France}
\and \author{W. W. Zeilinger\altaffilmark{}}         
\affil{Institut f\" ur Astronomie der Universit\" at  Wien\\
 T\" urkenschanzstra$\ss$e 17\\
 A-1180 Wien, Austria}

\begin{abstract}
We detected recent star formation in nearby early-type galaxies located in  
low density environments, with {\it GALEX} Ultraviolet (UV) imaging. 
Signatures of star formation may be present in the nucleus and 
in outer rings/arm like structures.  Our  study suggests that 
such star formation may be induced by different triggering mechanisms,
such as the inner secular evolution driven by bars, and minor accretion phenomena.   
We investigate the nature of the (FUV-NUV) color vs. Mg$_2$
correlation, and suggest that it relates to
``downsizing'' in galaxy formation. 
\end{abstract}


\section{Introduction}

Although early-type galaxies (ETGs hereafter) are considered the fossil record 
of the process of galaxy evolution, there is growing evidence that ``rejuvenation''
episodes may occur in their star formation history. 
In this context, the halo mass of ETGs is the main  driver for their evolution,
however also the environment plays a significant role \citep{Clemens06,Clemens09}.
According to these studies, the galactic nuclei (SDSS fibers sample the  
central 3\arcsec) of ETGs in low density environments (LDE hereafter)
are about 20\% younger than those of their cluster counterparts. Wide-field (1.25 degrees 
diameter), deep far-UV (1344 $-$ 1786 \AA) and near UV (1771 $-$ 2831 \AA) 
imaging from {\it GALEX} (see for details Bianchi these proceedings) 
is greatly contributing to this view. Using {\it GALEX}
data, e.g. \citet{Schawinski07} found that 30\% of massive ETGs
show ongoing star formation, and that this fraction is higher
in LDE  than in the high-density environments.

 \citet{Rampazzo07} and \citet{Marino09} showed that 
ETGs with shell structures (indicative, according to simulations,
 of recent accretion episodes) host a ``rejuvenated'' nucleus
 \citep[see also][for optical spectroscopic studies]{Longhetti00}.
Similar results have been obtained by \cite{Jeong09} for the {\tt SAURON}
 galaxy sample \citep{deZeeuw02}.

We are performing a comprehensive, multi wavelength
study of 65 nearby ETGs,  a large fraction of which show
 ionized gas emission,  predominantly located in LDE 
\citep[][A07 and A10 hereafter]{Annibali07,Annibali10}.
The sample is composed of 70\% of elliptical and 30\%
of S0s \citep[see both][]{RC3,RSA}. Anyway, from the kinematic
point of view,  about 68\% of our ETGs have fast rotator 
characteristics in the $\epsilon$ vs. $V/\sigma_e$
plane (see Appendix A in A10).

Here we present the GALEX view  for a sub-sample 
of 40 ETGs, out of  the 65 in the original sample, available
in the NASA archive. The UV spectral region is sensitive to 
even small amounts of  recent star formation, and is thus 
effective in unveiling  possible ``rejuvenation'' episodes.
For a detailed presentation of the GALEX FUV and NUV 
photometry we refer to  \citet{Marino10}.
We review our multi-wavelength approach in quest of
signatures of recent star formation in ETGs.
Finally we combine the analysis of the UV photometry with 
optical line-strength indices and show preliminary results.

\begin{figure}
\includegraphics[width=6.7cm]{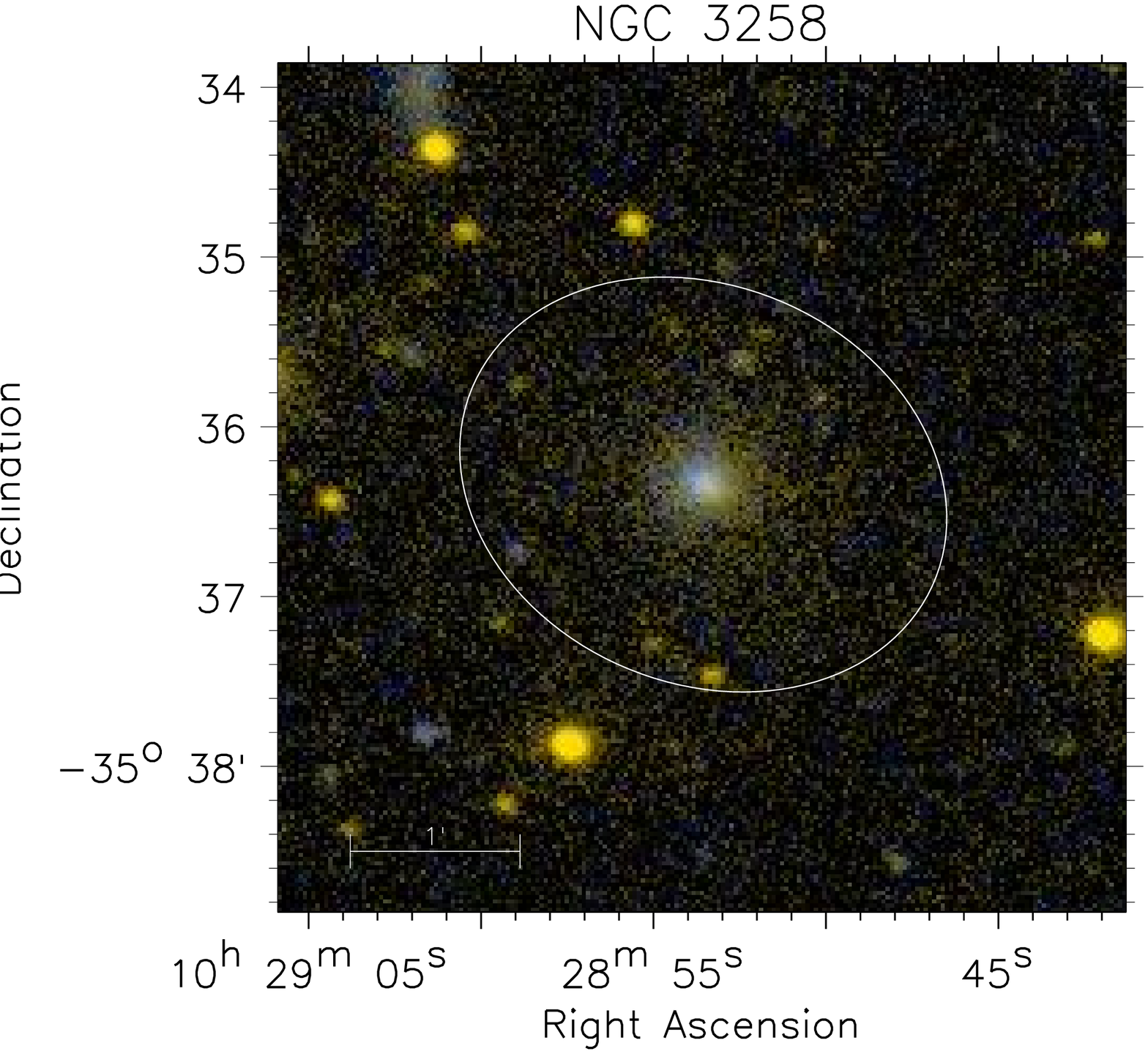}
\includegraphics[width=6.7cm]{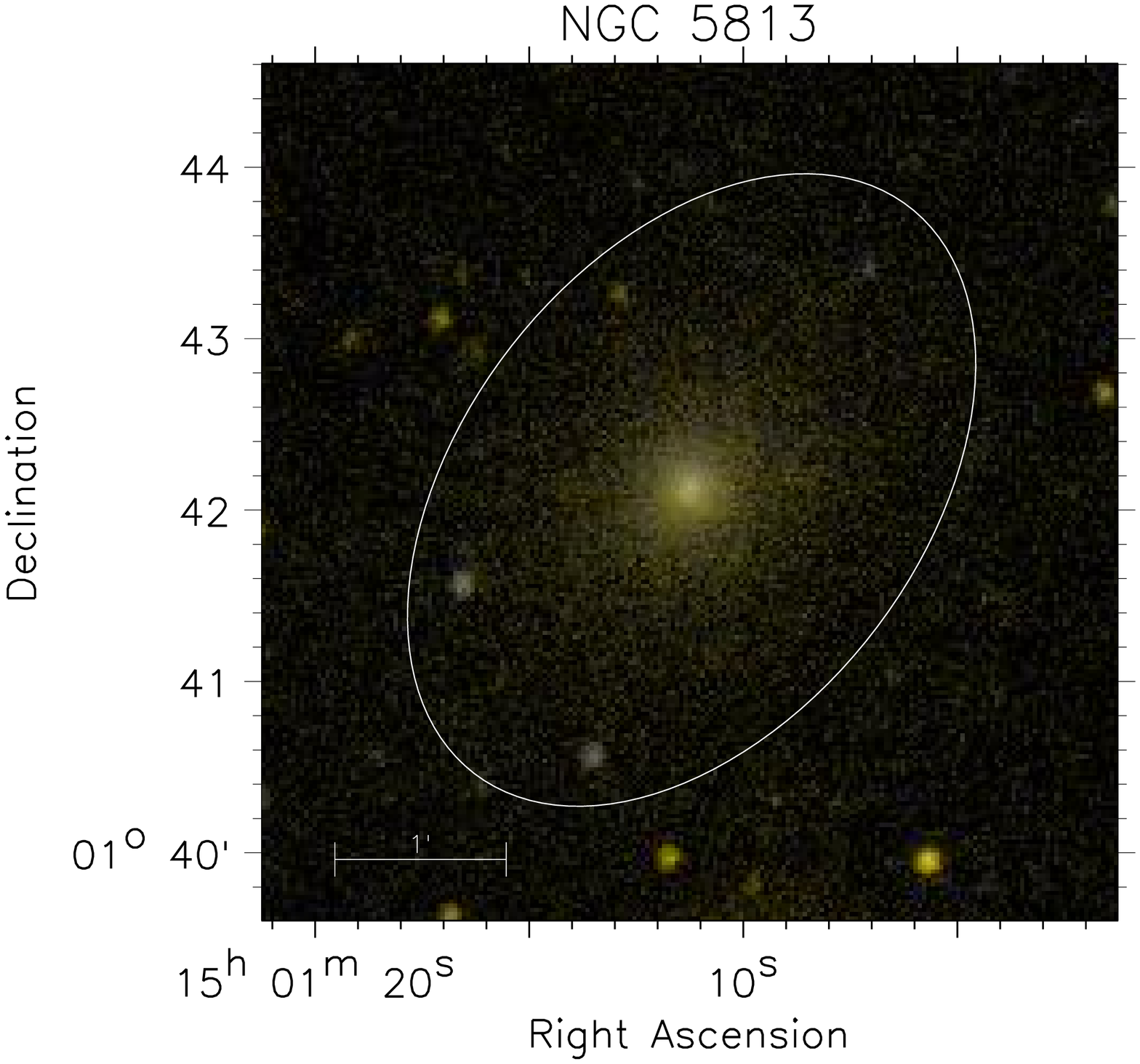}
\includegraphics[width=6.7cm]{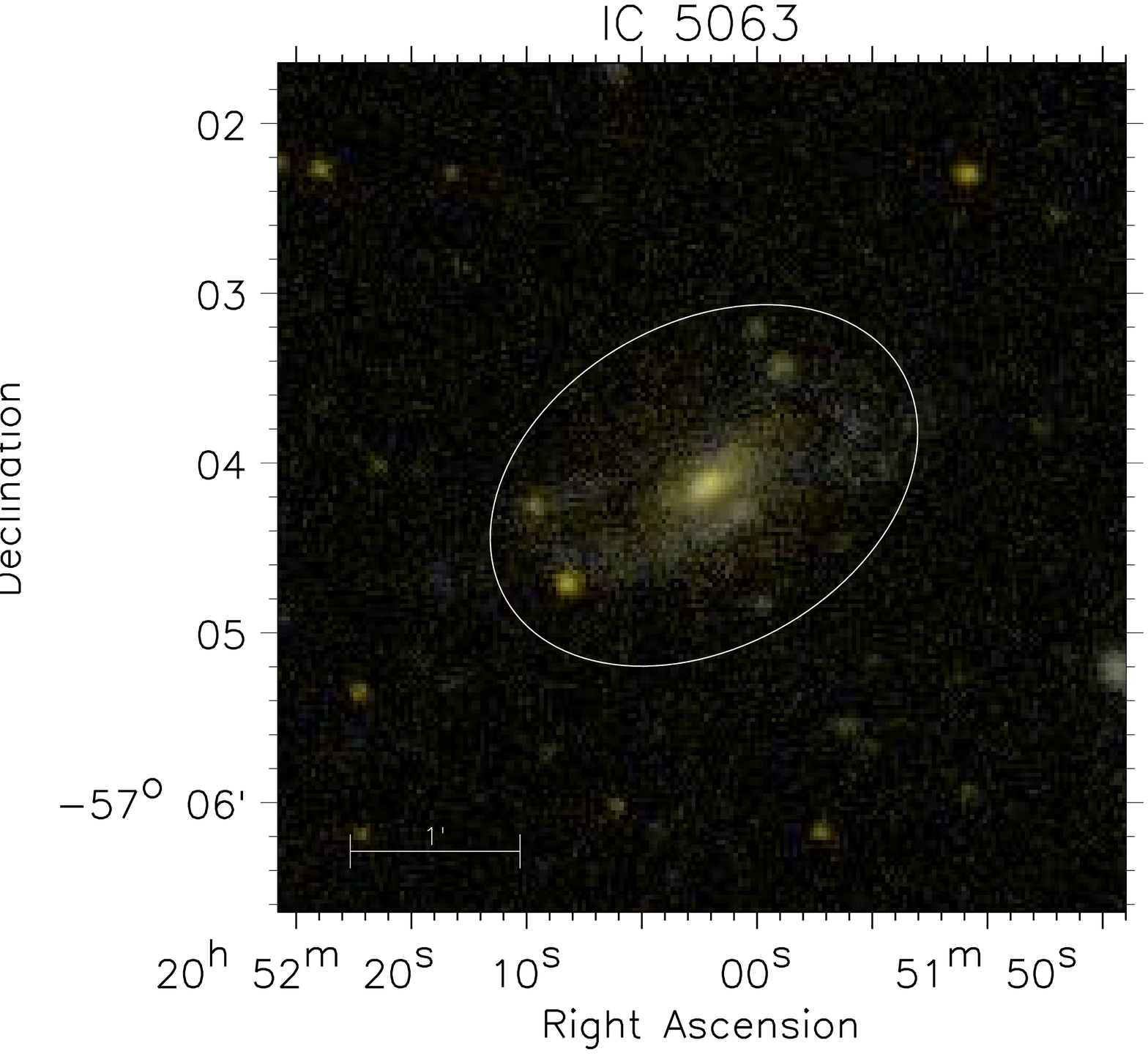}
\caption{GALEX false color image (FUV blue; NUV yellow) of NGC 3258, NGC 5813
and of the Seyfert~2 IC~5063 (adapted from \citet{Marino10}). The ellipse marks the isophote
at $\mu_B=25$~mag~arcsec$^{-2}$ (D$_{25}$).  Optical line-strength
indices indicate that NGC~3258 has a luminosity weighted age of
4.5$\pm$0.8 Gyr, while NGC 5813 is an old ETG with an
age of 11.7$\pm$1.6 \citep[see][]{Annibali07}} 
\label{fig1}
\end{figure}

\begin{figure}
\begin{center}
\includegraphics[width=6.2cm]{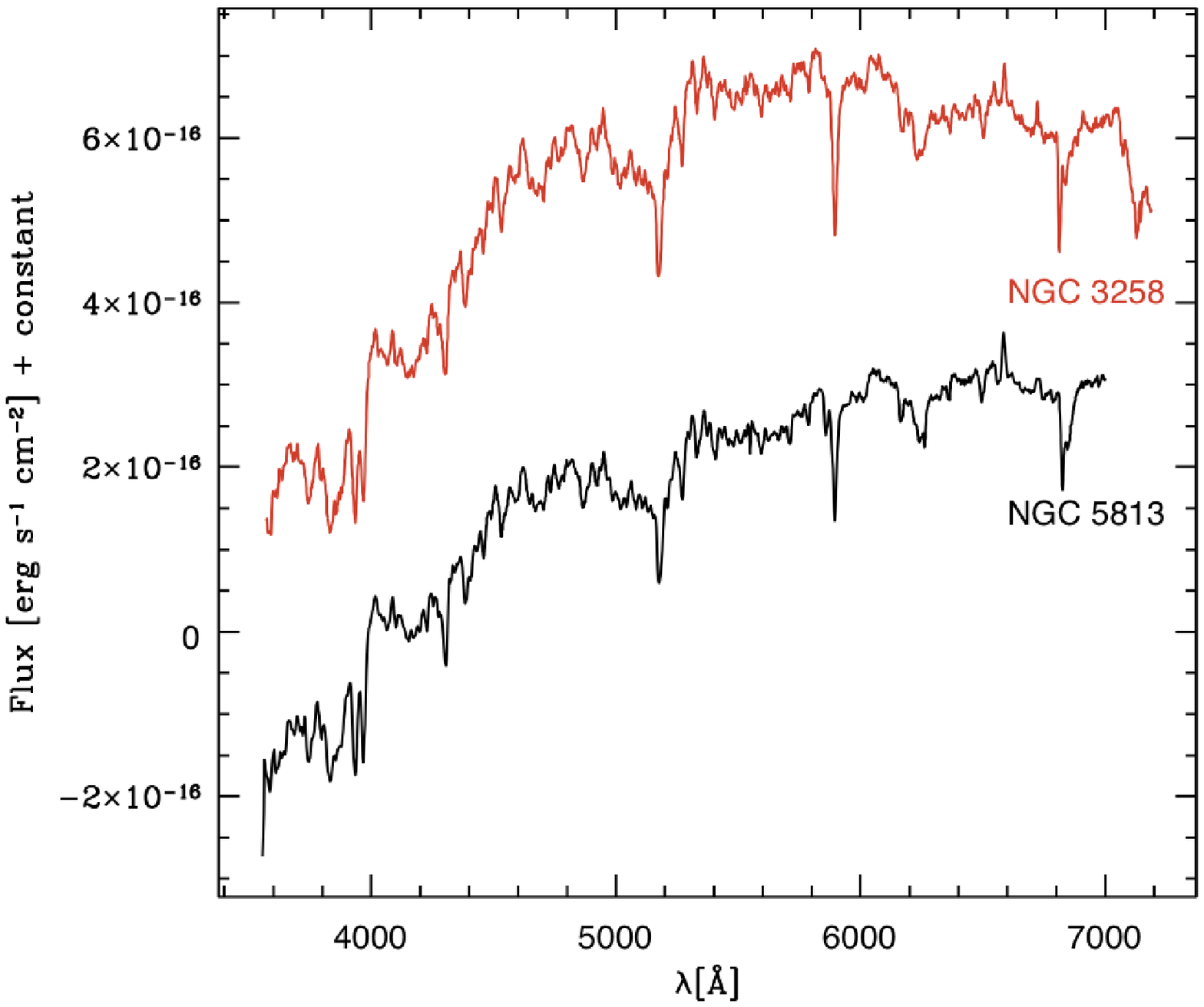}
\includegraphics[width=6.0cm]{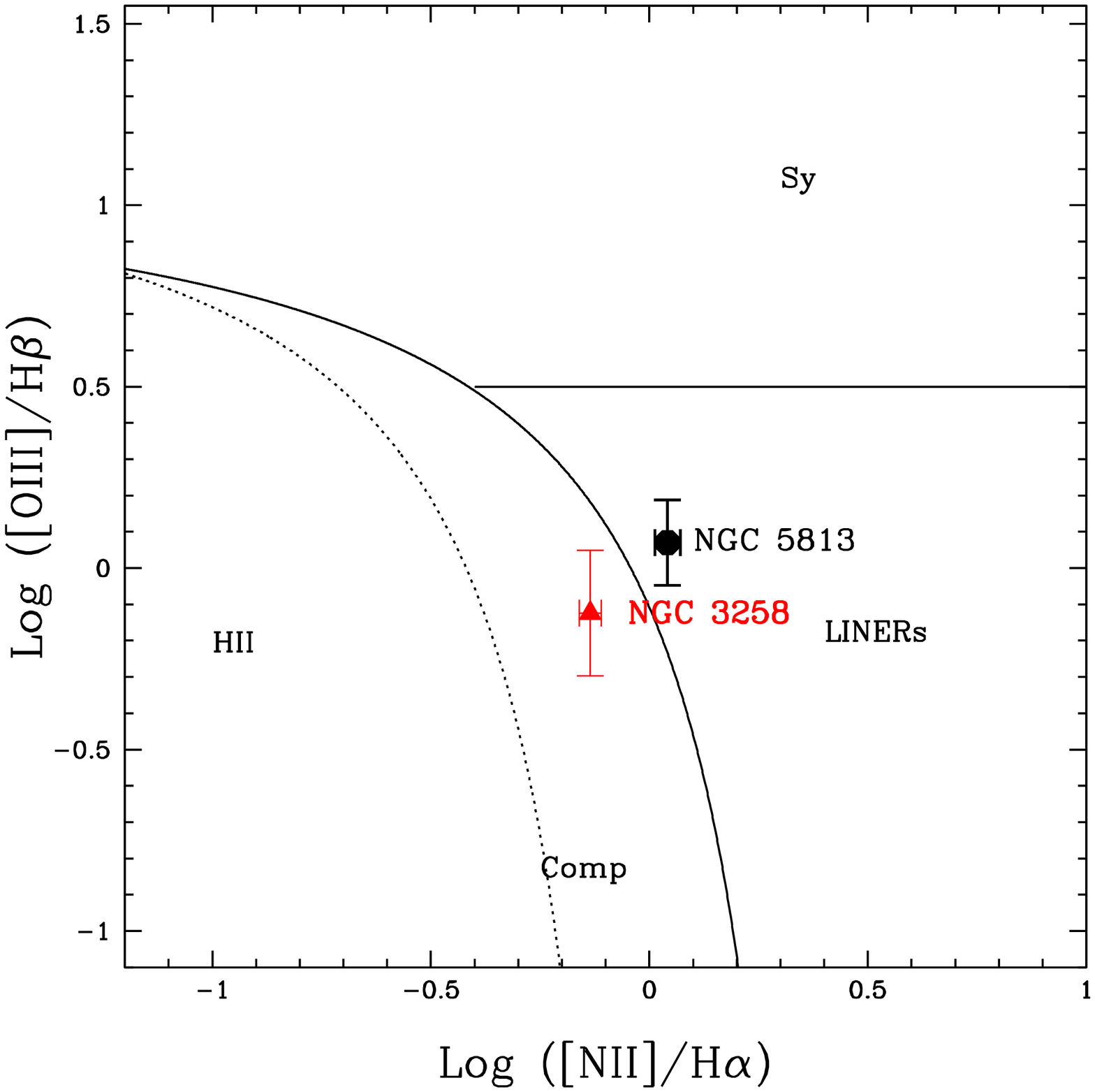}
\includegraphics[width=6.0cm]{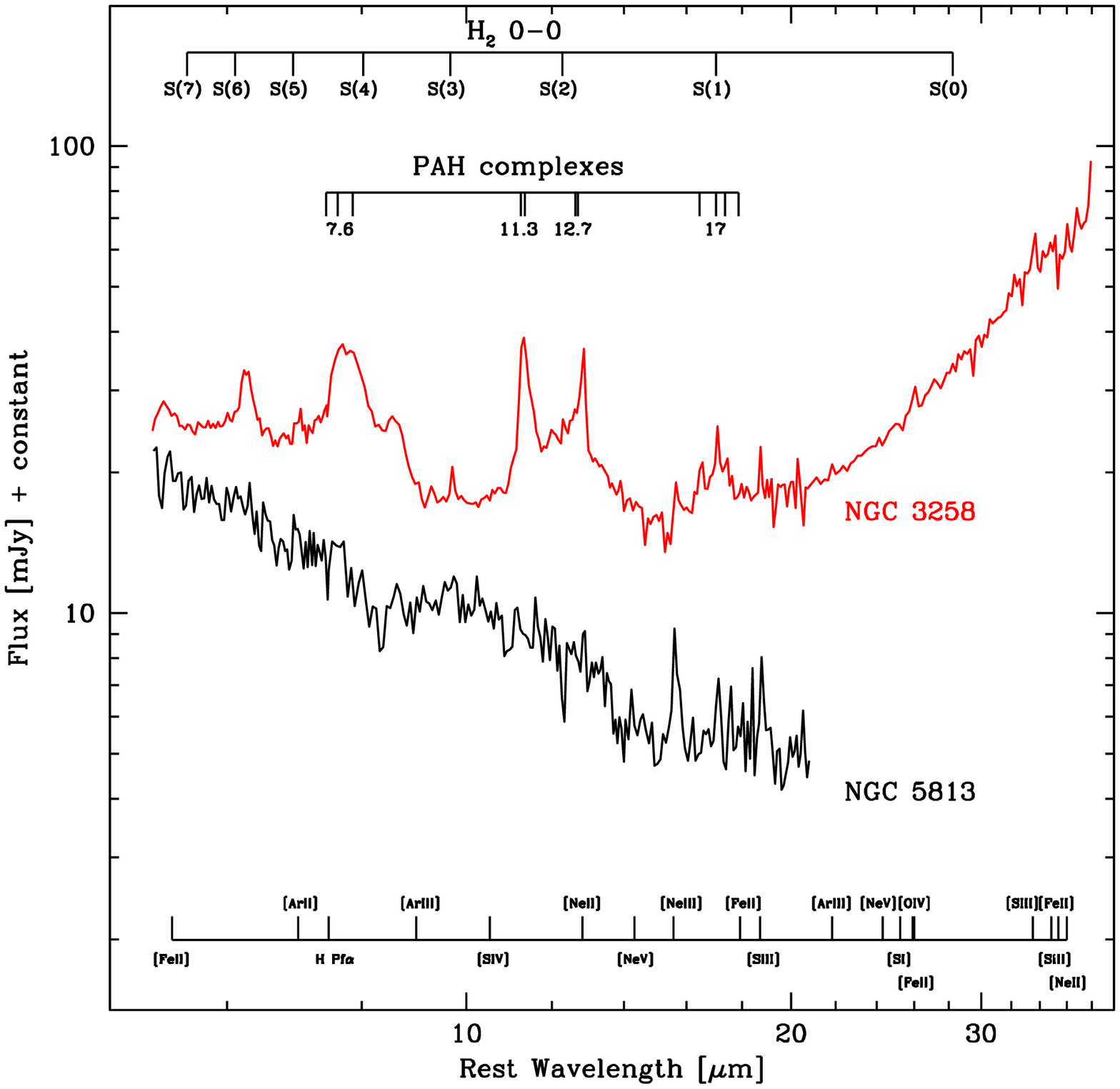}
\caption{Optical \citep[see details in][]{Rampazzo05,Annibali06} and MIR 
spectra of NGC~3258 and NGC~5813 galaxies imaged in Figure~1.
The top panel shows the nuclear spectra of the two galaxies,
fluxes are shifted arbitrarily, in the range 3750 $\leq \lambda \leq$ 7250 \AA.
In the mid panel we show the position of the two galaxies in the
typical BTP diagram (see for details A10) which indicates
 the LINER/Composite nature of the two galaxy nuclei. 
 The bottom panel shows
 the {\it Spitzer}-IRS MIR  spectra. Notice the presence of PAHs in 
 the spectrum of NGC 3258 at odds with NGC 5813. A cold dust 
 component is also visible in the NGC 3258 spectrum
} 
\end{center}
\label{fig2}
\end{figure}

\begin{figure}
\includegraphics[width=\columnwidth]{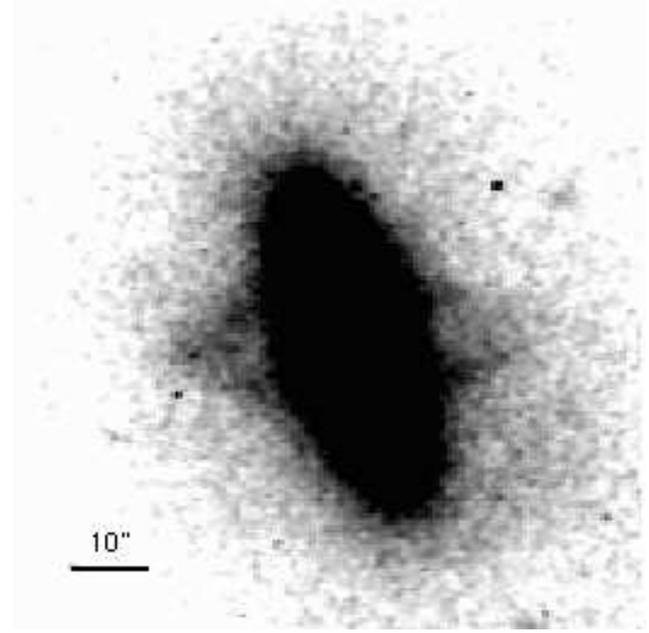}
\includegraphics[width=\columnwidth]{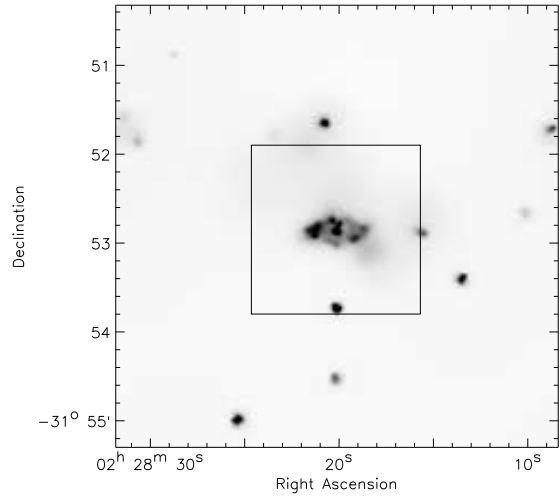}
\caption{ B-band (top panel) and {\it GALEX}
FUV  images (bottom panel) of MCG-05-07-001, a southern
polar ring galaxy.  In the FUV band  (the square encloses the field of view 
of the B-band image), only the ring and the nucleus of the galaxy 
are visible: the old main body of the galaxy disappears
 \citep[see][for details]{Marino09}} 
\label{fig3}
\end{figure}

\begin{figure}
\includegraphics[width=\columnwidth]{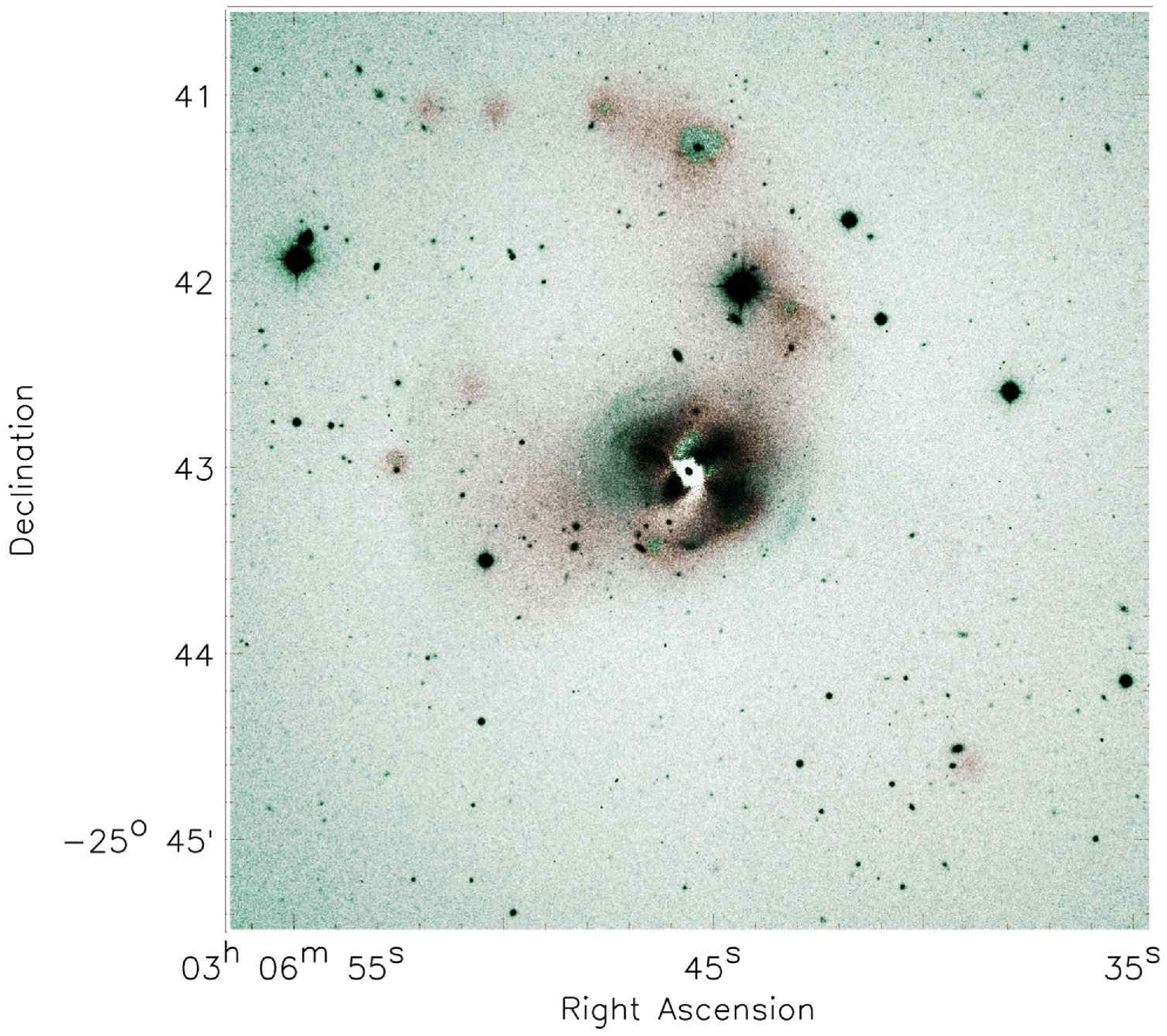}
\caption{ The {\it GALEX} FUV emission (red areas) is superposed 
to the residual of the optical image of NGC~1210 after  the main 
body of the galaxy has been subtacted. The residuals show a wide system 
of shells and a complex central fine structure. A ring of HI emission,
which perfectly superposes to the {\it GALEX} FUV, has been detected by
\citet{Schminovic01} \citep[see][for details]{Marino09}  } 
\label{fig4}
\end{figure}

\section{Looking for signatures of recent star formation: a multi-wavelength sweep}

In Figure~\ref{fig1} we show the composite {\it GALEX}
 image of three ETGs in the \citet{Marino10} sample, NGC~3258, NGC~5813
 and IC~5063.

 Notwithstanding the unperturbed morphology revealed by both 
 the deep optical \citep{Tal09} and our UV imaging,
 NGC~3258 and NGC 5813 hide a different recent star formation
 history.  The analysis of the optical line-strength indices 
 performed by A07 shows that  NGC~3258 has a ``young'' 
 average stellar population (luminosity-weighted age of 
 4.5$\pm$0.8 Gyr). 
 At odds,  NGC~5813 is  a relatively old ETG (11.7$\pm$1.6 Gyr). 
The nuclear optical spectra of these galaxies are shown in the
top panel of Figure~2. Like the imaging,  their spectra 
 do not show peculiarities: emission lines are, indeed, found in a 
 relatively large fraction  of ETGs (between 45\% to 80\% depending 
 on the sample analyzed in the literature, see e.g. the
 review in A10).   The classic [NII]($\lambda$ 6584)/H$\alpha$ vs. 
 [OIII]($\lambda$ 5007)/H$\beta$ diagnostic diagram 
 \citep[][BPT diagram]{Baldwin81}  is shown in the mid panel 
 of Figure~2.  The diagram shows  that the two galaxies host a 
 LINER/Composite nucleus (A10) as large fraction of ETGs 
 \citep[see e.g.][]{Phi86,Sarzi06}.

 The different star formation history suggested by the optical results 
 is supported  by our {\it Spitzer} mid-infrared (MIR)
 observations \citep{Panuzzo10}. Despite the remarkable similarity  
 of the shape of their optical spectra, the nuclear MIR emission  
 of NGC 3258 and NGC 5813, shown in the
 bottom panel of Figure~\ref{fig2}, is completely different.
 The Polycyclic Aromatic Hydrocarbons emission (PAH)  complexes  at 
 6.2 $\mu$m, 7.7 $\mu$m,  11.3 $\mu$m and 17 $\mu$m dominate the 
 MIR spectrum of  NGC 3258. Their  ratios indicate that the galaxy nucleus 
 has undergone a recent starburst \citep[see details in][]{Panuzzo10,Vega10}. 
At odds, the MIR spectrum of NGC 5813 displays a broad emission feature
around 10 $\mu$m attributed to the silicate emission arising from 
dusty circum-stellar envelopes of  O-rich AGB stars,  superimposed on the
photospheric stellar continuum from red giant stars. This is the typical spectrum of
passively evolving ETGs \citep[see e.g.][]{Bressan06}.

IC~5063, at the bottom of Figure~\ref{fig1}, is a well known Seyfert 2 \citep[see][]{Veron06}.
It was the first galaxy in which a fast neutral hydrogen outflow was discovered  
\citep[see][and reference therein]{Morganti07}.  Morganti et al. have also found 
evidence of cold and warm gas outflows and noticed that the outflow rate, 
associated to the HI,  is in the same range as for ``mild''
starburst driven superwinds in ULIRGs. Although limited to the inner kpc region
the observed outflows may have sufficient kinetic power to have a significant impact
on the evolution of the interstellar medium of IC~5063.  
In A10 we show that in IC~5063 a Seyfert like ionization extends 
out to our  largest sample radius, r$_e$/2, i.e. about 3 kpc 
(H$_0$=75 km s$^{-1}$ Mpc$^{-1}$). In general in A10
we show that a link exists between the AGN phenomenon 
and the presence of recent star formation in the galaxy nucleus.
With the {\it GALEX}  imaging we extend our search for star formation
to the galaxy outskirts.  Figure~\ref{fig1} shows the complex, outer, blue 
ring-like structure detected in UV in IC~5063.  
The origin of such  structure  may be connected with an accretion
episode as the complex dust-lane crossing the galaxy. 

\begin{figure}[]
\includegraphics[width=\columnwidth]{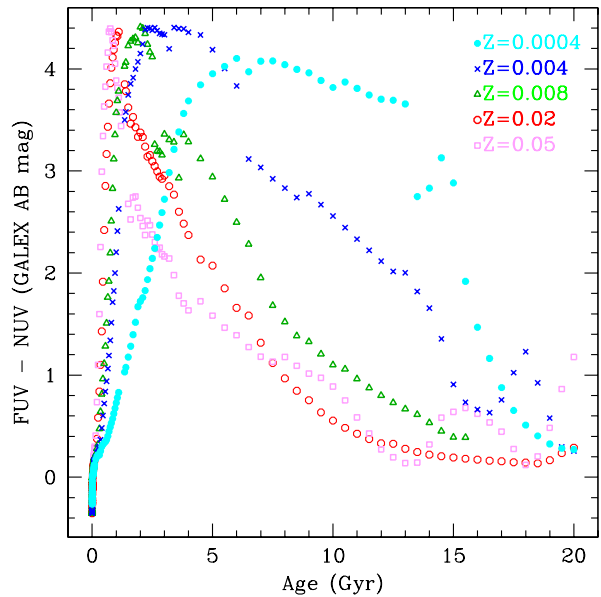}
\caption{SSP (FUV$-$NUV) GALEX color in the AB mag 
system as a function of age, for five metallicity values
(Z$_{\odot}=$0.0156, \citet{Caffau09})} 
\label{fig5}
\end{figure}

\begin{figure}[tb]
\includegraphics[width=\columnwidth]{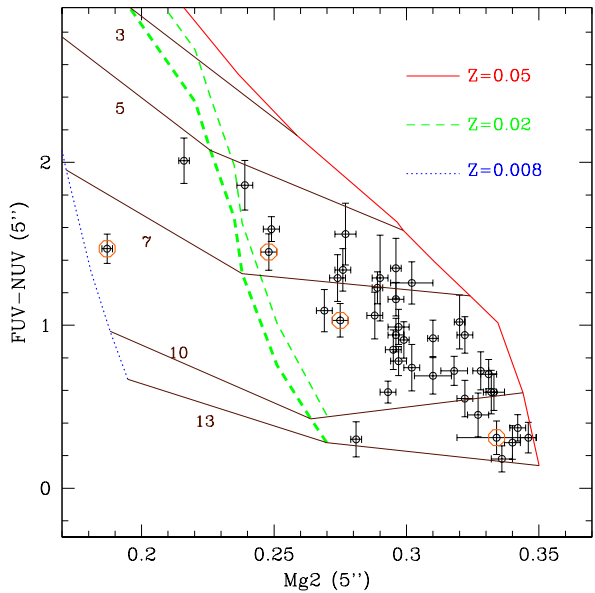}
\caption{(FUV$-$NUV) colors versus the Mg$_2$ index measured 
within a 5'' radius aperture \citep{Marino10}.
The big circles indicate galaxies classified as Seyferts by \citet{Annibali10}.
Overplotted are SSP models for different ages (13, 10, 7, 5 and 3 Gyr) and metallicities
(Z$=$0.05, 0.02, 0.008), and $E(B-V)=0$. The thin dashed line is for 
a Z$=$0.02 SSP with $E(B-V)=0.3$} 
\label{fig6}
\end{figure}

In our sample,  four barred S0 galaxies, 
namely NGC~1533, NGC~2962, NGC~2974 and NGC~3489, show outer 
blue ring/arm-like structures. The recent star formation detected in
this kind of rings is discussed in the Marino et al. paper in this conference
proceedings.  We emphasize here that such outer ring/arm-like 
structures are likely due to inner secular evolution of a galaxy
driven by the bar \citep[see for a kinematical review][]{Moiseev09}. 
The bar formation may have been induced by a galaxy-galaxy interaction 
within the groups \citep[see e.g. pioneering simulations of][]{Noguchi90}.

The rings  in the above barred S0s have a different origin from 
polar rings and/or collisional rings and tidal structures seen 
in some ETGs where accretion or major merger episodes 
have recently occurred.  In Figure~\ref{fig3} we show an example
of a polar-ring galaxy  studied in \citet{Marino09}. A very
young, lower than 1 Gyr, stellar population is  present in the ring
and is evidence of a recent accretion/merging phenomenon. ETGs with
polar rings in the Local Universe are quite rare: only 0.5\% of  ETGs 
show such phenomenon \citep{Whitmore90}. Rings
closely associated with a bar are much more common, they are
present in about 20-30\% of lenticular and spiral galaxies
\citep{Buta96}.

Seven ETGs in the \citet{Marino10} sample show a shell
system, namely NGC 1553, NGC 2974, NGC 4552, NGC 6958,
NGC 7135, NGC 7192 and IC 1459 \citep[see the compilation of][]{MC83,Tal09}.
Shells or ripples are revealed in about 15-20\% of 
ETGs \cite[see e.g.][]{MC83,Reduzzi96}.
The most successful  models for explaining shells 
consider the accretion of  a smaller companion by a giant ETG
\citep[see e.g.][]{Quinn84,Dupraz86,Ebrova09}. The small
galaxy accreted disintegrates in the giant potential and its
stars begin to oscillate forming shells. If the accretion is ``wet'', 
stars and gas segregates. The gas dissipates the orbital kinetic energy
and falls toward the primary center settling into compact disks or rings
in/around the primary nucleus \citep{Weil93}. NGC 1210 is a beautiful example 
of shell galaxy.  In Figure~\ref{fig4},  the {\it GALEX} FUV flux distribution  
provides clear evidence for an accretion event. 
\citet[][]{Marino09} suggest that also the nucleus of NGC ~1210 has
been rejuvenated and the accretion episode, that triggered the
star formation, should be dated between 2 and 4 Gyr ago.

\section{The nuclear (FUV-NUV) vs. Mg$_2$  relation}

Both the secular galaxy evolution and the accretion driven
evolution tend to refuel the galaxy nucleus of fresh gas
and trigger star formation. 

Complementing UV colors with optical line strength indices 
is a way to characterize the star formation 
history of ETGs, and  in particular to unveil the possible 
presence of secondary star formation events.

Combining  {\it GALEX}  photometry with optical line-strength 
indices, \cite{Donas07} found a tight anti-correlation  
 between the (FUV$-$NUV) color and the Mg$_2$ index for 
 a sample of 130 nearby ETGs \citep[see also][]{Rampazzo07},
 in the sense that the (FUV$-$NUV) color becomes bluer 
 as the Mg$_2$ index increases.
The same trend appears between the UV color and the galaxy central 
velocity dispersion. \cite{Donas07} suggest that these correlations are 
mainly driven by metallicity and reflect blanketing in the NUV being 
correlated with the overall metallicity.

To understand how the (FUV$-$NUV) color is affected by age and metallicity, 
we computed synthetic magnitudes in the {\it GALEX} systems for a set 
Simple Stellar Populations (SSPs) of different ages and metallicities 
(Bressan, unpublished,  see also \cite{Clemens09,Chav09,Bianchi09}).
Our results are illustrated in Figure~\ref{fig5}. 

At the youngest ages, the SSPs have the bluest (FUV$-$NUV) colors
($\sim$ $-$0.1 mag for a 10 Myr old population with Z$=$0.02),
as the O-type and B-type stars contribute most of the FUV flux;
then they become redder as the age increases (up to $\sim$ 4.4 mag). For ages older 
than 1 to a few Gyrs (depending on metallicity), the trend is inverted,
and the (FUV$-$NUV) color becomes progressively bluer as
the age increases ($\approx$ 0.3 mag for a 13 Gyr old population 
with Z$=$0.02).

This behavior can be understood considering that the NUV band is highly 
dominated by turnoff stars. 
On the other hand, the FUV emission of old ``normal'' stellar 
populations is dominated by post asymptotic giant branch (PAGB) stars.
In younger and more metal poor populations, the turnoff is bluer 
and more luminous, and emits more in the NUV; at the same time, the 
contribution to the FUV from PAGB stars diminishes because, 
in spite of the higher luminosity, the duration of the 
PAGB phase gets much shorter (i.e., the fuel decreases). 
This implies that for ages older than 1-5 Gyrs (depending on the metallicity)
 the (FUV$-$NUV) color at any given age is redder for lower metallicity. 
However, the situation could be more complex if  ``anomalous'' sub-populations, 
such as those suggested to be responsible for the UV-upturn of elliptical galaxies, 
are present.  For populations younger than 1-5 Gyr, 
the turnoff starts to contribute significantly to the  
FUV band, and the reddening of  the (FUV$-$NUV) color
with age is modulated by metallicity.
It follows that, for populations older than $\sim$ 1 to a few Gyrs,
bluer (FUV$-$NUV) colors can be caused by either
older ages or larger metallicities. 

Our analysis aims at clarifying  whether  the observed  
(FUV$-$NUV) vs $Mg_2$ (or $\sigma$) is driven 
by age or by metallicity (or both).
In Figure~\ref{fig6} we show  the  (FUV$-$NUV) color measured
in the central 5\arcsec\ of the galaxy versus the Mg$_2$ index 
for our sample of 40 ETGs. Superimposed are our SSP models.  
Lines of constant age run almost horizontal, 
while lines of constant metallicity are more vertical. 
Figure~\ref{fig6} shows that the slope of the (FUV$-$NUV) vs. Mg$_2$ correlation 
is significantly steeper than the lines of constant age and  it is very close 
to the slope defined by the models of constant metallicity. 
This suggests that the  (FUV$-$NUV) vs. Mg$_2$ trend is driven more by age than by metallicity, 
with redder ETGs being also younger. 
In this sense, the (FUV$-$NUV) vs. Mg$_2$ anti-correlation is more an aspect of 
``downsizing'' rather than of the color-metallicity relation.

There is a caveat on the above conclusions due to the fact that 
the plotted models do not account for the effect of $\alpha$-enhanced 
chemical compositions. However, the effect of the $\alpha$-enhancement 
on the Mg$_2$  index turns out less important than the dependence 
on the total metallicity. For example, for a 7 Gyr old population, a factor 
of 2 increase in Z ($+$0.3 dex) implies a $\sim$ 20 \% increase in 
Mg$_2$ index, while a 0.3 dex increase in [$\alpha$/Fe] implies 
a 7\% increase in Mg$_2$ index. This is because the [$\alpha$/Fe] 
enhancement is mainly due to an iron depletion rather than to an 
Mg enhancement. On the other hand, we cannot quantify with the 
present models the effect of the [$\alpha$/Fe] enhancement on the UV colors.

The possible contribution of ``anomalous'' sub-populations, 
such as those suggested to be responsible for the UV-upturn of elliptical galaxies, 
and the [$\alpha$/Fe] enhancement on the UV colors will be the subject of a 
forthcoming paper.

New UV observational facilities, like the WSO-UV observatory \citep{Shustov09,Shustov11},
will be crucial to fully understand the evolution of ETGs.

\acknowledgments

We acknowledge financial contribution from the agreement 
ASI$-$INAF I/009/10/0.
AM acknowledges support from the Italian Scientist and Scholar 
of North America Foundation (ISSNAF) through an ISSNAF fellowship 
in Space Physics and Engineering, sponsored by Thales Alenia Space.
Based on Galaxy Evolution Explorer  {\it GALEX} observations: GI3-0087 PI R. Rampazzo 
and from archival data. {\it GALEX}  is operated for NASA by
California Institute of Technology under NASA contract NAS-98034.
{\it GALEX} is a NASA Small Explorer, launched in April 2003. 

\nocite{*}
\bibliographystyle{spr-mp-nameyear-cnd}

\end{document}